\documentclass[aps,prb,twocolumn,superscriptaddress,showpacs]{revtex4}
\usepackage{graphicx}
\usepackage{bm}
\usepackage{epsfig}
\usepackage{ulem}
\usepackage{color}
\usepackage{mathrsfs}
\usepackage{dcolumn}
\usepackage{setspace}
\usepackage{array}
\usepackage{amsmath}
\usepackage{amssymb}
\usepackage{gensymb}
\usepackage{bibentry,natbib}
\usepackage{booktabs}
\begin{document}
\bibliographystyle{unsrt}

\title{Chiral Wave-packet Scattering in Weyl Semimetals}

\author{Qing-Dong Jiang}
\affiliation{International Center for Quantum Materials, School of
Physics, Peking University, Beijing 100871, P.R. China}
\author{Hua Jiang}
\affiliation{College of Physics, Optoelectronics and Energy, Soochow University, Suzhou 215006, P.R. China}
\author{Haiwen Liu}
\affiliation{International Center for Quantum Materials, School of Physics, Peking University, Beijing 100871, P.R. China}
\affiliation{Collaborative Innovation Center of Quantum Matter, Beijing 100871, P.R. China}
\affiliation{Center for Advanced Quantum Studies, Department of Physics, Beijing Normal University, Beijing 100875, P.R. China}
\author{Qing-Feng Sun}
\affiliation{International Center for Quantum Materials, School of Physics, Peking University, Beijing 100871, P.R. China}
\affiliation{Collaborative Innovation Center of Quantum Matter, Beijing 100871, P.R. China}
\author{X. C. Xie}
\affiliation{International Center for Quantum Materials, School of Physics, Peking University, Beijing 100871, P.R. China}
\affiliation{Collaborative Innovation Center of Quantum Matter, Beijing 100871, P.R. China}
\begin{abstract}
In quantum mechanics, a particle is best described by the wave packet instead of the plane wave. Here, we study the wave-packet scattering problem in Weyl semimetals with the low-energy Weyl fermions of different chiralities. Our results show that the wave packet acquires a chirality-protected shift in the single impurity scattering process. More importantly, the chirality-protected shift can lead to an anomalous scattering probability, thus, affects the transport properties in Weyl semimetals. We find that the ratio between the transport lifetime and the quantum lifetime increases sharply when the Fermi energy approaches the Weyl nodes,  providing an explanation of the experimentally observed ultrahigh mobility in topological (Weyl or Dirac) semimetals.
\end{abstract}
\maketitle

\section{ Introduction}

With the discovery of two-dimensional ultra-relativistic material---graphene, many exotic phenomena have found their wonderland, and can further be realized in simple table-top experiments.\cite{neto} Beyond graphene, three-dimensional ultra-relativistic materials, dubbed Dirac semimetals, have been predicted and confirmed recently.\cite{zwang,liang,liu,neupane,jeon,jkim,SBorisenko,lphe} The energy dispersion is linear near the band touching points (Dirac points), which is protected by the crystalline symmetry. When either the time-reversal symmetry or the inversion symmetry is broken, the Dirac semimetals (DSMs) evolve into the Weyl semimetals (WSMs), in which low energy Weyl fermions are embedded.\cite{xwan,balents,gxu1,burkov,hosur2012charge,luweylpoints,syyang1,Hosur-Qi,shuang,Weng} Both DSMs and WSMs are called topological semimetals (TSMs). Like the DSMs, WSMs also have linear energy dispersion near the band touching nodes (Weyl nodes). Importantly, the Weyl nodes can be viewed as magnetic monopoles in k-space, and generate the Berry curvature $\boldsymbol \Omega$.\cite{balents} Due to the transverse anomalous velocity $\bold v_A=\boldsymbol \Omega\times e\bold E$,\cite{xiao2010berry,chang1996berry} a host of novel transport phenomena can happen in WSMs, such as the chiral magnetic effect,  the topological Imbert-Fedorov effect, etc.\cite{nielsen,son,dtson,jychen,mstone,cduval,zyuzin,cxliu,qdjiang,chiralyang,ybaum}. The Weyl semimetal state has been confirmed by photoemission and magneto-transport experiments in the noncentrosymmetric TaAs family of compounds.\cite{Lv,Xu,Lu,Potter,shekhar,xiong,huang,zhang,Yang}

All of the transport experiments confirmed two striking universal features of the TSMs: (i) the negative magnetoresistance when the magnetic field is parallel to the electric field,\cite{huang,zhang,xiong,hli} and (ii) the ultrahigh mobility in the TSMs systems.\cite{liang,huang,zhang,Yang,Ghimire,shekhar,mnali,zzhu,tafti} The negative magnetoresistance feature has been predicted theoretically, which is caused by the chiral anomaly effect in WSMs. However, the physical mechanism that leads to the ultrahigh mobility remains puzzling. The mobility is determined by the transport lifetime $\tau_{t}$. Conventionally, $\tau_t$ is regarded as the same order as the quantum lifetime (or single particle scattering time) $\tau_{q}$.\cite{dsarma,MacLeod} In the case of two-dimensional electron gas (2DEG) in GaAs/AlGaAs, the ratio $R_\tau=\tau_t/\tau_q$ can be large ($10-100$) because the scattering centers are separated from the carriers.\cite{dsarma,mapaalanen,jpharrang,ptcoleridge} The charge fluctuations in the dopant layer lead only to small-angle scatterings, which strongly limit $\tau_q$ but hardly affect $\tau_t$.\cite{ptcoleridge} Unexpectedly, a recent experiment showed that the ratio $R_\tau$ can be as large as $10^4$ in TSM.\cite{liang} If $\tau_q$ is assumed to be of normal magnitude, large ratio $R_\tau$ implies ultrahigh mobility in TSM. However, there is no obvious separation of the scattering centers from the conduction electrons in TSM, yet $R_\tau$ is much larger. There should exist an unknown mechanism that strongly suppresses the backscattering.

In this paper, we address the unknown mechanism that leads to the ultrahigh mobility in the TSMs. First, we show that a chirality-protected shift occurs in the wave-packet scattering process. Our result shows that the chirality-protected shift reaches its maximum in the backscattering case.  Second, we discuss how the shift modifies the effective impact parameter of the wave packet, and thereby results in the anomalous scattering probability. Third, to validate the argument we made, we present a full quantum calculation on the wave-packet scattering process, and  the results are consistent with our semiclassical argument. At last, we calculate the transport lifetime $\tau_t$ and the quantum lifetime $\tau_q$. Our results show that the ratio $R_\tau=\tau_t/\tau_q$ increases steeply as the Fermi energy approaches the Weyl nodes.
\begin{figure}[!htb]
\centering
\includegraphics[height=6.2cm, width=8.5cm, angle=0]{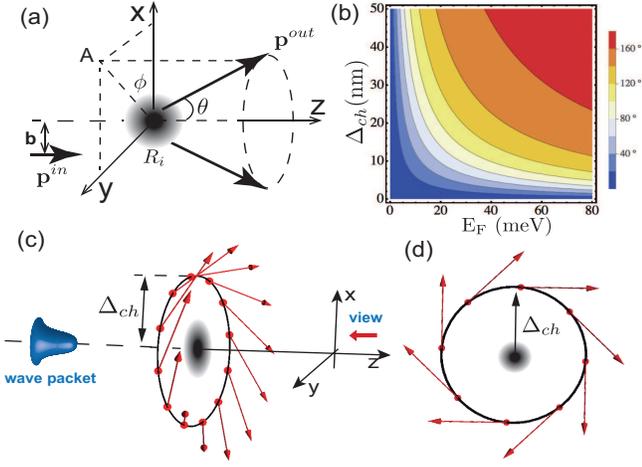}
\caption{The chirality-protected transverse shift occurs in the scattering process in WSMs.
(a) The schematic of the wave-packet scattering. $\bold b$ is the impact parameter; $\theta$ and $\phi$ are two angles that characterize the outgoing direction of the scattered wave. (b) The chirality-protected transverse shift vs Fermi energy for different scattering angles.  (c) The chirality-protected shift $\Delta_{ch}$ emerges in the wave-packet scattering process. The red arrows stand for the outgoing direction of the scattered wave($\mathcal C=-1$). (d) shows the momentum texture of the scattered wave ($\mathcal C=-1$). When one views the scattering process in the z direction, the momentum direction of the outgoing wave forms a texture, and the radius of the texture is the chirality-protected shift.  For Weyl fermions with chirality $\mathcal C=1$, the radius remains the same, whereas the momentum texture reverses. \label{fig:1}}
\end{figure}

\section{The chirality-protected transverse shift}
In this section, we derive the chirality-protected transverse shift in the wave-packet scattering process. The effective Hamiltonian of Weyl fermions is\cite{Hosur-Qi}
\begin{eqnarray}
\mathcal H=\sum_{i,j\in{x,y,x}} \hbar v_{ij}\,k_i\sigma_j,
\end{eqnarray}
where $v_{ij}$ have the dimensions of velocity, $k_i$ is the wave vector, and $\boldsymbol \sigma=(\sigma_x, \sigma_y, \sigma_z)$ are the Pauli matrices. The chirality (handedness) of Weyl fermions is defined as $\mathcal C=\rm sgn\left[det(v_{ij})\right]$.  For convenience, we study the isotropic Weyl semimetal described by the Hamiltonian $\mathcal H=\mathcal C ~\hbar v \bold k\cdot\boldsymbol \sigma$. We consider that the wave packet is scattered by a single impurity $R_i$ in WSM [Fig. 1(a)]. In Ref.\cite{qdjiang}, the authors find that a topological transverse shift occurs in the reflection process in WSMs. Therefore, we also expect that a transverse shift may appear in the impurity scattering process. In principle, there can also be a longitudinal shift in the wave-packet scattering process. However, the longitudinal shift should not contribute to the total angular momentum, and we do not consider it here. Because of the rotational symmetry about the  z axis, the z component of the total angular momentum should be conserved in the wave-packet scattering process, by which the transverse shift can be obtained. Because of the massless nature of Weyl fermions, the chirality is the same as the helicity, which can be understood as the projection of the spin angular momentum to the canonical momentum.
If the chirality $\mathcal C=1(-1)$, that means the spin direction is (anti)parallel to the momentum. For the Weyl fermions with the chirality $\mathcal C=\pm 1$, the spin angular momentum changes $\Delta \bold S=\pm\frac{\hbar}{2}(\hat r-\hat e_z)$, where $\hat r=\left(\rm sin\theta cos\phi, sin\theta sin\phi, cos\theta\right)$ is the unit vector in the same direction with the outgoing wave packet. The orbital momentum changes $\Delta \bold L=\boldsymbol \Delta_{ch}\times \bold p^{out}$, where $\boldsymbol \Delta_{ch}=\Delta_{ch}\left(\rm sin\theta_0cos\phi_0, sin\theta_0sin\phi_0,cos\theta_0\right)$ is the transverse shift, and $\bold p^{out}=p\,\hat r$ is the momentum of the scattered wave packet. $\phi_0$ and $\theta_0$ characterize the direction of the transverse shift. We initially assume that the chirality-protected shift is perpendicular to incident momentum $\bold p^{in}$ and out scattered momentum $\bold p^{out}$. Thus, one can obtain $\theta_0=\pi/2$ and $\phi_0=\phi\pm\pi/2$. The total angular momentum conservation in the z direction gives
$\Delta S_z+\Delta L_z=\frac{\hbar}{2}(\rm cos\theta-1)\pm\Delta_{ch}\times p\,sin\theta$.
Therefore, one can readily obtain the absolute value of the transverse shift
\begin{eqnarray}
\rm{\Delta_{ch}=\frac{1}{2k}tan\frac{\theta}{2}}.
\end{eqnarray}
Because of the rotational symmetry, the absolute value is independent of the azimuthal angle $\phi$. Fig. 1(b) shows the chirality-protected shift versus the Fermi energy with different scattering angles. The direction of the shift $\boldsymbol \Delta_{ch}$ can also be obtained by considering the conservation of the total angular momentum. For Weyl fermions with chirality $\mathcal C=1$, the spin angular momentum decreases in the scattering process, leading to increase of the orbital angular momentum. Therefore, we obtain $\phi_0=\phi-\frac{\pi}{2}$ for $\mathcal C=1$. Analogously, we  obtain $\phi_0=\phi+\frac{\pi}{2}$ for $\mathcal C=-1$. Therefore, the transverse shift is
\begin{eqnarray}
\rm{\boldsymbol\Delta_{ch}=\frac{1}{2k}tan\frac{\theta}{2}\left[cos(\phi-\delta),sin(\phi-\delta)\right]},
\end{eqnarray}
where $\delta=\pm \pi/2$ for Weyl fermions with chirality $\mathcal C=\pm 1$, respectively. The phase shift $\delta=\pm \pi/2$ characterizes the chiral nature of the wave-packet scattering process.
Fig. 1(c) and (d) show the scattering momentum texture due to the transverse nature of the  chirality-protected shift.

\section{Anomalous scattering probability}
In the wave-packet scattering process, one can define the impact parameter $\bold b$, which is the perpendicular distance between the wave-packet center and the target center(Fig. 1(a)).\cite{cjjoachain} According to the wave-packet scattering theory, the scattering probability $P(\theta,\bold b)$ is exponentially decayed with the increase of the impact parameter $\bold b$, i.e. $P(\theta,\bold b)\propto e^{-\bold b^2/\Delta_r^2}\sigma(\theta)$, where $\Delta_r$ is the width of the wave packet and $\sigma(\theta)$ is the scattering cross section.\cite{cjjoachain}
Since we have shown that there is a chirality-protected transverse shift in the scattering process, the effective impact parameter $\bold{\tilde{b}}$ should be renormalized due to this transverse shift, i.e.,
\begin{eqnarray}
\bold{\tilde{b}}=\bold b+\boldsymbol \Delta_{ch}.
\end{eqnarray}
Therefore, because of the transverse shift $\boldsymbol \Delta_{ch}$, the scattering probability should be
\begin{eqnarray}
P(\theta, \bold b)\propto e^{-\bold{\tilde{b}}^2/\Delta_r^2}\sigma_W(\theta),
\end{eqnarray}
where $\sigma_W(\theta)$ is the scattering cross section for Weyl fermions.
As is shown in Fig. 1(b), the shift $\Delta_{ch}\rightarrow \infty$  in the backscattering process ($\theta=\pi$), which indicates that the backscattering is strongly suppressed by the factor $e^{-\bold{\tilde{b}}^2/\Delta_r^2}$.
Even in non-backscattering case ($\theta\neq\pi$), $\Delta_{ch}\rightarrow \infty$ when the Fermi energy approaches to the Weyl nodes ($E_F\rightarrow 0$). This indicates that only small-angle scatterings are possible in the wave-packet scattering process in WSMs. We remark that the physical mechanism of the ultrahigh mobility in topological semimetals is different from that in GaAs-based 2DEG.\cite{dsarma,mapaalanen,jpharrang,ptcoleridge}
In the appendices, we also calculate the wave-packet scattering for the Dirac fermions in graphene.  The scattering probability is still exponentially decayed with the increase of the impact parameter $\bold b$, i.e., $P(\theta, \bold b)\propto e^{-\bold b^2/\Delta_r^2}\sigma_D(\theta)$, where  $\sigma_D(\theta)$ is the scattering cross section for Dirac fermions.  Notably, there is no chirality-protected shift $\Delta_{ch}$ for the Dirac fermions in graphene.

\section{Quantum calculation on the wave-packet scattering probability}
Previously, we discussed that the chirality-protected shift could lead to the anomalous scattering probability. To validate this argument, we perform the full quantum calculation on the wave-packet scattering problem. Notably, the two valleys in Weyl semimetals are usually separated with a large momentum difference. In our paper, we use the Born approximation to deal with the Coulombic impurity scattering problem. The inter-node scattering requires large momentum transfer, which will seldom happen due to the long range feature of the Coulomb potential. As a contrast, we firstly calculate the scattering cross section for the plane-wave scattering process in WSMs. Assume that the plane wave is incident in the z direction, which is expressed as $\psi^{in}=(1,0)^Te^{ikz}$.
The incident plane wave is scattered by a single impurity with the potential $V(\bold r)$. Using the first Born approximation, the outgoing wave function is
\begin{eqnarray}
\begin{split}
\psi^{out}(\bold r)=&\psi^{in}(\bold r)+\int d\bold r^{\prime}G(\bold r-\bold r^{\prime})V(\bold r^{\prime})\psi^{in}(\bold r^{\prime})\\
=&\psi^{in}(\bold r)+f(\theta,\phi)\frac{e^{ikr}}{r},
\end{split}
\end{eqnarray}
where $G(\bold r)$ is the Green function of the Weyl Hamiltonian $\mathcal H$, and $f(\theta,\phi)$ is the scattering amplitude.
The differential scattering cross section is
\begin{eqnarray}
\sigma(\theta,\phi)=|f|^2=2\left(\frac{\hbar v k}{4\pi \hbar^2 v^2}\right)^2~|M_0|^2\rm(1+cos\theta).
\end{eqnarray}
Here, $M_0=\int d\bold r e^{-i\bold q\cdot \bold r}V(\bold r)$, where $\bold q=k\hat r-k \hat e_z$ denotes the transfer of the wave vector in the scattering process. The angle $\theta$ is the scattering angle satisfying ${\rm{cos\theta}}=r_z/r$.

Next, we consider the wave-packet scattering process. The wave packet is described by the Gaussian distribution $\varphi(\bold k)=\left(\frac{1}{\pi\Delta_k^2}\right)^\frac{3}{4}e^{-\frac{(\bold k-\bold k_0)^2}{2\Delta_k^2}}$, where $\Delta_k$ is the width of the wave packet in k space, and $\bold k_0$ is the mean wave vector.\cite{cjjoachain} We assume that the wave packet is incident in the z direction, and is scattered by an impurity $R_i$ located at the origin of the coordinate system. The general eigenfunction of the Hamiltonian [Eq.(1)] is $\rm\left(cos\frac{\theta_0}{2}, sin\frac{\theta_0}{2}e^{i\phi_0}\right)^T$, with the angles $\theta_0={\rm arccos}\frac{k_z}{k}$ and $\phi_0={\rm arcsin}\frac{k_y}{\sqrt{k_x^2+k_y^2}}$ characterizing the propagating direction of the plane wave. Therefore, the incident wave packet can be expressed as
\begin{eqnarray}
\begin{split}
\psi_g^{in}(\bold r)=&\int\frac{d^3k}{(2\pi)^\frac{3}{2}}\varphi(\bold k)e^{-i\bold k\cdot(\bold r_0-\bold r)-iEt}~
\left(\begin{array}{cc} \rm cos\frac{\theta_0}{2}\\ \rm  sin\frac{\theta_0}{2}e^{i\phi_0}\end{array}\right)\\
=&\int\frac{d^3k}{(2\pi)^\frac{3}{2}}\varphi(\bold k)e^{-i\bold k\cdot(\bold r_0+\bold v_0 t-\bold r)}~
\left(\begin{array}{cc} \rm cos\frac{\theta_0}{2}\\ \rm  sin\frac{\theta_0}{2}e^{i\phi_0}\end{array}\right).
\end{split}
\end{eqnarray}
$\bold r_0=(r_{0x}, r_{0y}, r_{0z})$ is the initial position of the wave packet, and  $\bold v_0=v\hat k_0$ is the propagating velocity of the incident wave packet.
In deriving the above expression, we used the approximation $E=\sqrt{\hbar v(k_x^2+k_y^2+k_z^2)}\approx\hbar \bold v_0\cdot\bold k$ (see the appendices).
According to the Born approximation, the outgoing wave function is
\begin{eqnarray}\begin{split}
\psi_g^{out}(\bold r)&=\psi_g^{in}(\bold r)+\psi_g^{s}(\bold r)\\
&=\psi_g^{in}(\bold r)+\int\frac{d^3k}{(2\pi)^\frac{3}{2}}\varphi(\bold k)\times\\
\int d\bold r^{\prime}&G(\bold r-\bold r^{\prime})V(\bold r^{\prime})e^{-i\bold k\cdot(\bold r_0+\bold v_0 t-\bold r^{\prime})}
\left(\begin{array}{cc}
\rm cos\frac{\theta_0}{2}\\\rm sin\frac{\theta_0}{2}e^{i\phi_0}
\end{array}
\right).
\end{split}\end{eqnarray}
Here, $\psi_g^{s}(\bold r)$ is the scattered wave function, which is obtained after taking the integral on $\bold r^{\prime}$.
\begin{eqnarray}
\psi_g^{s}(\bold r)=
\frac{1}{r}\left(
\begin{array}{cc}
\rm 1+cos\theta, sin\theta e^{-i\phi}\\
\rm sin\theta e^{i\phi}, 1-cos\theta
\end{array}
\right)
\left(\begin{array}{cc}
M\\N
\end{array}\right),
\end{eqnarray}
where $M=-\frac{\hbar v k}{4\pi \hbar^2 v^2}M_0\Delta_k^{\frac{3}{2}}\pi^{-\frac{3}{4}}exp\left\{-\frac{(r-r_{0z}-v_0t)^2\Delta_k^2}{2}\right\}\times\\ exp\left\{-\frac{b^2\Delta_k^2}{2}\right\}e^{ik_0(r-r_{0z}-v_0t)}$ and $N=(-ir_{0x}\Delta_k^2+r_{0y}\Delta_k^2)\cdot\frac{1}{2k}M$. In the expression of $M$ and $N$, $\bold b=(b_x,b_y)=(r_{0x},r_{0y})$ and $b=\sqrt{b_x^2+b_y^2}$ represents the impact parameter. In the appendices, we present the full details of the derivation of the expression of $M$ and $N$. Let us assume the detector has an area given by $r^2 d\Omega$ perpendicular to the radial direction. The  particles flows through the detector with the velocity $v_0$. Therefore, in an infinitesimally small time interval $t$---$t+dt$, all of the probability contained in the volume $r^2 d\Omega\times v_0 dt$ flows through the detector.  Consequently, the probability of observing a scattered particle in the time interval $t$---$t+dt$ is $|\psi_g^s(\bold r)|^2~r^2d\Omega \times v_0dt$. Therefore, one can obtain the total probability of detecting a particle at $d\Omega$:
\begin{eqnarray}
\begin{split}
P(\theta,\phi,\bold b)=&\int_{-\infty}^{\infty}(r^2d\Omega\times v_0 dt)|\psi_g^{s}|^2\\
= &2(\frac{\hbar v k}{4\pi \hbar^2 v^2})^2\Delta_k^2\pi^{-1}|M_0|^2\rm  (1+cos\theta)\times\\
exp\{-\Delta_k^2&\rm \left[\left(b_x+\Delta_{ch}\cdot sin\phi\right)^2+\left(b_y-\Delta_{ch}\cdot cos\phi\right)^2\right]\},
\end{split}
\end{eqnarray}
where $\Delta_{ch}=\frac{1}{2k}{\rm tan\frac{\theta}{2}}$. Strikingly, we find that $\Delta_{ch}$ is exactly the chirality-protected shift we previously obtained by the angular momentum conservation method. This result strongly supports the validity of our argument on the wave-packet scattering, i.e, the chirality-protected transverse shift modifies the impact parameter effectively, and thereby leads to the anomalous scattering probability of the wave packet.

\section{Quantum lifetime vs. transport lifetime}
The quantum lifetime and transport lifetime are two important time scales in condensed matter physics. The quantum lifetime $\tau_q$ measures the inter-collision events in all directions, which can be obtained from the Shubnikov-de Haas (SdH) oscillations. In contrast, the transport lifetime $\tau_{t}$ measures the collision events in one particular direction, which can be obtained from the conductivity.\cite{MacLeod}
Conventionally, the quantum lifetime and the transport lifetime are usually of the same order. However, in Ref.\cite{liang}, the authors find that the ratio between the transport lifetime and quantum lifetime can be as large as $\rm{10^4}$ in TSMs. Indeed, almost all of the transport measurements of the TSMs show the ultrahigh mobility, which challenges our consensus.\cite{tafti} In the TSMs, for instance, the ``ideal" $\rm Cd_3As_2$ lattice, there are 64 sites for $\rm Cd$ ions in each unit cell, and that  exactly $1/4$ of the sites are vacant, which can easily attract impurities and act as the scattering centers.\cite{liang} This means that even in the ``ideal"  $\rm Cd_3As_2$ material, the mobility should be severely suppressed by the presence of the vacancies. Remarkably, the puzzled ultrahigh mobility can be explained in the wave-packet scattering scenario. For example, in the case of zero impact parameter($\bold b=0$), the chirality-protected shift $\Delta_{ch}$ reaches its maximum leading to the infinitely large effective impact parameter $\tilde b$, which strongly suppresses the backscattering.
In the plane-wave scattering scenario, the quantum lifetime is obtained from the scattering cross section, i.e. $1/\tau_{q0}=v_f n_i\int \sigma(\theta)d\theta d\phi$, where $v_f$ is the Fermi velocity, $n_i$ is the impurity concentration. By contrast, the transport lifetime is $1/\tau_{t0}=v_f n_i\int\sigma(\theta){\rm (1-cos\theta)}d\theta d\phi$. $R_{\tau0}=\tau_{t0}/\tau_{q0}$ is the ratio between the transport lifetime and quantum lifetime in the plane-wave scattering process. In analogy with the lifetime defined above, we can define the quantum lifetime and transport lifetime in the wave-packet scattering process. In the wave-packet scattering process, the scattering cross section is replaced by the scattering probability, i.e. $\sigma(\theta,\phi)\longmapsto \int_0^{b_c} d\bold b P(\theta,\phi,\bold b)$, where $b_c$ is the cutoff of the impact parameter decided by the density of the impurities (see the appendices).  Therefore, one can define the quantum lifetime and transport lifetime:
\begin{eqnarray}
\frac{1}{\tau_q}&=&v_f n_i\int_0^{b_c} d\bold b\int d\theta d\phi P(\theta,\phi,\bold b)\\
\frac{1}{\tau_{t}}&=&v_f n_i\int_0^{b_c}d\bold b \int d\theta d\phi {\rm \left(1-cos\theta\right)}P(\theta,\phi,\bold b)
\end{eqnarray}
The ratio between the transport lifetime and the quantum lifetime is $R_{\tau}=\tau_{t}/\tau_q$. We calculate the quantum lifetime, transport lifetime and the ratio $R_\tau$.
Assume the density of the impurities is about $10^{18}~\rm cm^{-3}$. To ensure the single impurity scattering process, we set the impact parameter cutoff as $b_c=5~\rm nm$, which is the half of the average distance between the two nearest impurities.
Fig. 2(a) shows the transport lifetime and quantum lifetime, which saturate at large Fermi energy. Fig. 2(b) shows the ratio $R_\tau$ versus the Fermi energy for wave-packet scattering and plane-wave scattering, respectively. In Fig. 2(b), the ratio $R_\tau$ increases steeply when the Fermi energy approaching to the Weyl nodes, and is saturated at large Fermi energy.\cite{footnote2} The large ratio $R_\tau$ indicates that the impurities can strongly limit $\tau_q$ but hardly affect $\tau_t$, which is the physical mechanism of the ultrahigh mobility in TSMs.

\begin{figure}[!htb]
\centering
\includegraphics[height=4.cm, width=8.4cm, angle=0]{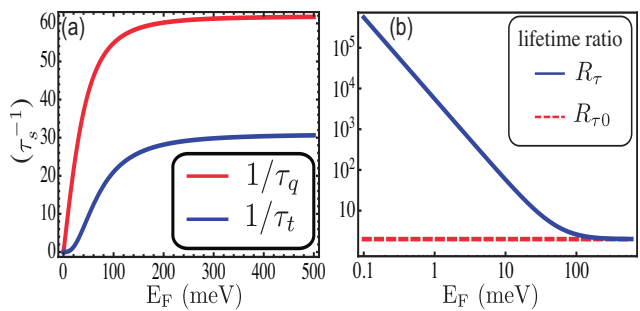}
\caption{Quantum lifetime and transport lifetime in WSMs.
(a) The quantum lifetime $\tau_q$ and transport lifetime $\tau_{t}$ vs the Fermi energy $E_F$ in WSMs. $\tau_s^{-1}=2\pi^{-1}v_f n_i|\frac{\hbar vk}{4\pi\hbar^2v^2}|^2|M_0|^2$ is the unit of the y axis. (b) The blue solid line is the lifetime ratio $R_\tau=\tau_t/\tau_q$ in the wave-packet scattering process, whereas the red dashed line is the lifetime ratio $R_{\tau 0}$ obtained from the plane-wave scattering.\label{fig:2}}
\end{figure}

In experiment, the exact impurity density is not easy to identify, but can be quantitatively identified. Therefore, we calculate how the impurity density $n_s$ influences the quantum lifetime $\tau_q$ and transport lifetime $\tau_{tr}$. Since the cutoff of the impact parameter is decided by the impurity density, the variation of the impurity density affects both the quantum lifetime and the transport lifetime. Fig. 3 shows how the quantum lifetime and the transport lifetime change with the impurity density. The Fermi energy $E_F$ is set as $10~\rm meV$. Fig. 3 implies that both the quantum lifetime and transport lifetime are strongly suppressed in the scenario of the wave-packet scattering. 

\begin{figure}[!htb]
\centering
\includegraphics[height=5.cm, width=7.cm, angle=0]{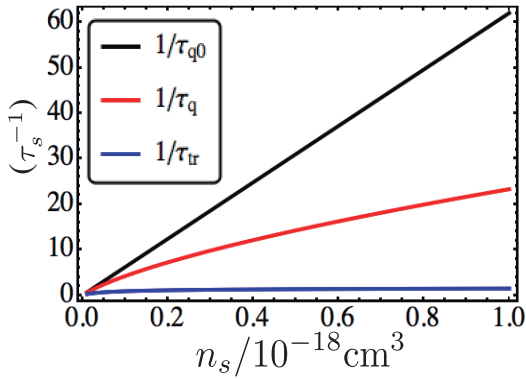}
\caption{Quantum lifetime and transport lifetime vs. impurity density in WSMs.
$\tau_{q0}$ is the quantum lifetime in the scenario of the plane-wave scattering. $\tau_q$ and $\tau_t$ are the quantum lifetime and transport lifetime in the scenario of the wave-packet scattering.\label{fig:3}}
\end{figure}

\section{Discussion}
Before the conclusion, we make several remarks on our work in this section.
(i) Side jump effect: The chirality-protected shift is reminiscent of the side jump effect in spin-orbit coupling systems.\cite{lberger,nnagaosa,jsinova,jychen2} However, the chirality-protected shift is very nontrivial because of the divergence of the Berry curvature near the Weyl points. In this paper, we study the impurity scattering effect of the Weyl fermions. In the backscattering case, the shift tends to infinity leading to vanished scattering probability, which is the reason for the ultrahigh mobility.
(ii) Wave-packet width: There are several important length scales, including the wave-packet width $\Delta_r$, impact parameter cutoff $b_c$, chirality-protected shift $\Delta_c$, and lattice constance $a$. Because we are using the continuous low-energy model, we demand that $\Delta_r>>a$. If we also demand that the final result does not depend on the wave-packet width, we require that $b_c>> \Delta_r$. In practice, we only need $b_c>1.5 \Delta_r$ to eliminate the wave-packet width $\Delta_r$ dependence of the final result. Therefore, in order to make our result solid, we need to combine the above conditions, i.e., $b_c>1.5\Delta_r>>a$.
(iii) Temperature influence: The real experiments are always done in finite temperature, and the temperature influence must be taken into account.  However, the temperature only serves as the fluctuation of the Fermi energy. When temperature is low enough, the final result will not change much. (iv) Large Fermi energy in Ref. [3]: We comment that our theory may not be applicable to Ref. [3], in which the authors report the observation of the ultrahigh mobility for large Fermi energy (232 $\rm meV$). In our model, the lifetime ratio $R_\tau$ is extremely large when Fermi energy approaches to the Weyl nodes, and can be strongly suppressed due to the increasing of the Fermi energy.\cite{clzhang2}


\section{Summary}
We found the chirality-protected shift in the wave-packet scattering process by using the angular momentum conservation method. The chirality-protected transverse shift modifies the effective impact parameters, and leads to the anomalous scattering probability. More importantly, we give a full quantum treatment of the wave-packet scattering process and derive the exact result of the wave-packet scattering probability. Our results show that the ratio between the transport lifetime and the quantum lifetime increases sharply when the Fermi energy approaches the Weyl nodes. The unexpected large ratio can lead to the experimental findings of the puzzling ultrahigh mobility in topological semimetals.

\begin{center}
\textbf{ACKNOWLEDGEMENTS}
\end{center}
This work was financially supported by NBRP of
China (Grants No. 2015CB921102, No. 2014CB920901,
No. 2012CB921303, and No. 2012CB821402); NSF China
under Grants No. 11534001, No. 11274364, No. 11374219,
and No. 11504008; and NSF of Jiangsu province (Grant
No. BK20130283). We also want to thank Shuang Jia and
Chenglong Zhang for the helpful discussion on their
experimental data \cite{clzhang2}\\

\begin{center}
\textbf{Appendix A: Detail Derivation of the Wave-packet Scattering for the Weyl Equation}
\end{center}

In this section, we present the systematic derivation of the wave-packet scattering in Weyl semimetals (WSMs), and compare it with the plane-wave scattering result.  First of all, we give the detail derivation for the plane-wave scattering.  Then, we construct the wave-packet scattering from the plane-wave scattering result.

\textit{Plane-wave scattering---}
The effective Hamiltonian of WSMs is
\begin{eqnarray}
\mathcal H=v\boldsymbol\sigma\cdot \bold p+V(\bold r),
\end{eqnarray}
where $V(\bold r)$ is the potential of the impurity.
First of all, we present a simple derivation of the Green function of the Weyl equation.
The Green function is defined as
\begin{eqnarray}
(v\boldsymbol\sigma\cdot \bold p-E)G(\bold r)=-\delta(\bold r)I_{2},
\end{eqnarray}
where $I_2$ is the unit $2\times2$ matrix. Although the Green function can be obtained systematically by using the  algebra relation $G(\bold r)=\left[E-\mathcal H\right]^{-1}$, we give a simpler derivation below.
Assume $G(\bold r)=\frac{1}{v^2\hbar^2}(v\boldsymbol\sigma\cdot\bold p+E)G_n(\bold r)$, then one can find
\begin{eqnarray}
\begin{split}
(v\boldsymbol\sigma\cdot \bold p-E)G(\bold r)=&\frac{1}{v^2\hbar^2}(v\boldsymbol\sigma\cdot \bold p-E)(v\boldsymbol\sigma\cdot \bold p+E)G_n(\bold r)\\
=&\frac{1}{v^2\hbar^2}(v^2p^2-E^2)G_n(\bold r)\\
=&-(\nabla^2+k^2)G_n(\bold r)=-\delta(\bold r).
\end{split}
\end{eqnarray}
We can see that the $G_n(\bold r)$ is exactly the Green function of Schr\"odinger equation, which can be find in text book: $G_n(\bold r)=-\frac{e^{ikr}}{4\pi r}$.\cite{jjsakurai} Then, one can obtain $G(\bold r)$ from the normal Green function $G_n(\bold r)$:
\begin{eqnarray}
\begin{split}
G(\bold r)=&\frac{1}{v^2\hbar^2}(v\boldsymbol\sigma\cdot\bold p+E)G_n(\bold r)\\
=&-\frac{\hbar v k}{4\pi \hbar^2v^2}
\left(\begin{array}{cc}
\rm cos\theta +1 & \rm sin\theta e^{-i\phi}\\
\rm sin\theta e^{i\phi} & \rm 1-cos\theta
\end{array}\right) \frac{e^{ikr}}{r},
\end{split}
\end{eqnarray}
where $\theta$ and $\phi$ characterize the outgoing angle of the scattering wave.
Assume the plane wave is incident in z-direction, i.e., $\psi^{in}=\left(\begin{array}{cc}1 \\0\end{array}\right)e^{ikz}$. Using the first order Born approximation, the scattered wave function is
\begin{eqnarray}
\begin{split}
\psi^s(\bold r)=&\int G(\bold r-\bold r^\prime)V(\bold r^{\prime})\psi^{in}(\bold r^\prime)d \bold r^{\prime}\\
=&-\frac{\hbar vk}{4\pi\hbar^2v^2} \left(\int d\bold re^{-i\bold q\cdot \bold r}V(\bold r)\right)\times \\
&\left(\begin{array}{cc}\rm 1+cos\theta & \rm sin\theta e^{-i\phi}\\
\rm sin\theta e^{i\phi} & \rm 1-cos\theta
\end{array}\right)\left(\begin{array}{cc}1\\
0
\end{array}\right)~\frac{e^{ikr}}{{r}}\\
=&f(\bold k,\hat r)~\frac{e^{ikr}}{{r}},
\end{split}
\end{eqnarray}
where $f(\bold k,\hat r)=-\frac{\hbar vk}{4\pi\hbar^2v^2} \left(\int d\bold re^{-i\bold q\cdot \bold r}V(\bold r)\right)\left(\begin{array}{cc}\rm 1+cos\theta \\
\rm sin\theta e^{i\phi}
\end{array}\right)$ is the scattering amplitude.
Therefore the differential scattering cross section can be obtained:
\begin{eqnarray}
\sigma(\Omega)=|f|^2~d\Omega=2(\frac{\hbar v k}{4\pi \hbar^2 v^2})^2~|M_0|^2(1+\rm cos\theta)~d\Omega,
\end{eqnarray}
where $\Omega$ is the solid angle and $M_0=\int d\bold re^{-i\bold q\cdot \bold r}V(\bold r)$.

\textit{Wave-packet scattering---}
In quantum mechanics, a real particle is more like a wave packet. Therefore, when considering a single particle scattering, wave-packet dynamic treatment is more reasonable.\cite{cjjoachain} We consider the wave packet described by the Gaussian distribution $\varphi(\bold k)=\left(\frac{1}{\pi\Delta_k^2}\right)^\frac{3}{4}e^{-\frac{(\bold k-\bold k_0)^2}{2\Delta_k^2}}$, where $\Delta_k$ is the width of the wave packet in k-space, and $\bold k_0$ is the mean wave vector.\cite{cjjoachain,ebeveren}
We assume that the wave packet is incident in z-direction, and is scattered by an impurity $R_i$ located at the origin of the coordinate system. The general eigenfunction of the Hamiltonian [Eq.(14)] is $\rm\left(cos\frac{\theta_0}{2}, sin\frac{\theta_0}{2}e^{i\phi_0}\right)^T$ with the angles $\theta_0={\rm arccos}\frac{k_z}{k}$ and $\phi_0={\rm arcsin}\frac{k_y}{\sqrt{k_x^2+k_y^2}}$ characterizing the propagating direction of the plane wave. Therefore, according to the linear superposition principle, the incident wave packet can be expressed as
\begin{eqnarray}
\begin{split}
\psi_g^{in}(\bold r)=&\int\frac{d^3k}{(2\pi)^\frac{3}{2}}\varphi(\bold k)e^{-i\bold k\cdot(\bold r_0-\bold r)-iEt}~
\left(\begin{array}{cc} \rm cos\frac{\theta_0}{2}\\ \rm  sin\frac{\theta_0}{2}e^{i\phi_0}\end{array}\right)\\
=&\int\frac{d^3k}{(2\pi)^\frac{3}{2}}\varphi(\bold k)e^{-i\bold k\cdot(\bold r_0+\bold v_0 t-\bold r)}~
\left(\begin{array}{cc} \rm cos\frac{\theta_0}{2}\\ \rm  sin\frac{\theta_0}{2}e^{i\phi_0}\end{array}\right).
\end{split}
\end{eqnarray}
$\bold r_0=(r_{0x}, r_{0y}, r_{0z})$ is the initial position of the wave packet, and  $\bold v_0=v\hat k_0$ is the propagating velocity of the incident wave packet.
In deriving the above expression, we used (i) the approximation $E=\sqrt{\hbar v(k_x^2+k_y^2+k_z^2)}\approx\hbar \bold v_0\cdot\bold k$; (ii) the approximation $kr\approx\bold k\cdot\hat k_0 r$. Here $k=\sqrt{k_x^2+k_y^2+k_z^2}$. The zeroth order term is $k=k_0=\sqrt{k_{x0}^2+k_{y0}^2+k_{z0}^2}$ and the first order term is $\frac{\partial k}{\partial k_x}|_{k_x=k_{x0}}(k_x-k_{x0})+\frac{\partial k}{\partial k_y}|_{k_y=k_{y0}}(k_y-k_{y0})+\frac{\partial k}{\partial k_z}|_{k_z=k_{z0}}(k_z-k_{z0})=(\bold k-\bold k_0)\cdot \hat k_0$. The zeroth order term plus the first order term gives $k\approx \bold k\cdot\hat k_0$.
According to the Born approximation, the outgoing wave function produced by the scattering of the wave packet is
\begin{eqnarray}\begin{split}
\psi_g^{out}(\bold r)&=\psi_g^{in}(\bold r)+\psi_g^{s}(\bold r)\\
&=\psi_g^{in}(\bold r)+\int\frac{d^3k}{(2\pi)^\frac{3}{2}}\varphi(\bold k)\times\\
&
\int d\bold r^{\prime}G(\bold r-\bold r^{\prime})V(\bold r^{\prime})e^{-i\bold k\cdot(\bold r_0+\bold v_0 t-\bold r^{\prime})}
\left(\begin{array}{cc}
\rm cos\frac{\theta_0}{2}\\\rm sin\frac{\theta_0}{2}e^{i\phi_0}
\end{array}
\right)\\
&=\psi_g^{in}(\bold r)-\frac{\hbar vk}{4\pi\hbar^2v^2}\left(\int d\bold rV(\bold r)e^{- i\bold q\cdot r}\right)\times \\
&\left(\begin{array}{cc}\rm 1+cos\theta & \rm sin\theta e^{-i\phi}\\
\rm sin\theta e^{i\phi} &\rm  1-cos\theta
\end{array}\right)\times\\
&\int \frac{d^3k}{(2\pi)^{3/2}}\varphi(\bold k)\frac{e^{i[kr-\bold k\cdot(\bold r_0+\bold v_0 t)]}}{{r}}\left(\begin{array}{cc}\rm cos\frac{\theta_0}{2}\\\rm sin\frac{\theta_0}{2}e^{i\phi_0}\end{array}\right).
\end{split}\end{eqnarray}
In the above formula, $\psi_g^{s}(\bold r)$ is the scattered wave function, which can be obtained after taking the integral on $\bold r^{\prime}$.

Now, one can expand the spinor of the incident wave function to the second order, i.e.,
\begin{eqnarray}
\left(\begin{array}{cc}cos\frac{\theta_0}{2}\\sin\frac{\theta_0}{2}e^{i\phi_0}\end{array}\right)\approx\left(\begin{array}{cc}1-\frac{1}{8}\left(\frac{k_r}{k_z}\right)^2\\ \frac{k_r}{2 k_z}e^{i\phi_0}\end{array}\right)\approx \left(\begin{array}{cc}1-\frac{1}{8}\frac{k_x^2+k_y^2}{k_0^2}\\\frac{k_x+ik_y}{2 k_0}\end{array}\right),
\end{eqnarray}
where $k_r=k sin\theta_0$, and $sin~\phi_0=k_y/k_r$.
To obtain the scattered wave function, one need to calculate the integral:
\begin{eqnarray}
\begin{split}
&\int \frac{d^3k}{(2\pi)^{3/2}}\varphi(\bold k)\frac{e^{i[kr-\bold k\cdot(\bold r_0+\bold v_0 t)]}}{{r}}
\left(\begin{array}{cc}1-\frac{1}{8}\frac{k_x^2+k_y^2}{k_0^2}\\\frac{k_x+ik_y}{2 k_0}\end{array}\right)\\
\approx&\frac{1}{r}\int \frac{d^3k}{(2\pi)^{3/2}}\left(\frac{1}{\pi\Delta_k^2}\right)^\frac{3}{4}e^{-\frac{[k_x^2+k_y^2+(k_z-k_0)^2]}{2\Delta_k^2}}\times\\
& e^{i\bold k\cdot(r\hat k_0- \bold r_0-\bold v_0 t)}
\left(\begin{array}{cc}1-\frac{1}{8}\frac{k_x^2+k_y^2}{k_0^2}\\\frac{k_x+ik_y}{2 k_0}\end{array}\right)\\
\end{split}
\end{eqnarray}
where $\bold b=(b_x,b_y)=(r_{0x},r_{0y})$ and $b=\sqrt{r_{0x}^2+r_{0y}^2}$ is the impact parameter. Now, the main task is to calculate the integral Eq.(23). We let
\begin{eqnarray}
\gamma_0=\int {d^3k}e^{-\frac{[k_x^2+k_y^2+(k_z-k_0)^2]}{2\Delta_k^2}}e^{i\bold k\cdot(r\hat k_0- \bold r_0 -\bold v_0 t)};
\end{eqnarray}
\begin{eqnarray}
\begin{split}
\gamma_1=&\int {d^3k}\left(\frac{k_x+ik_y}{2 k_0}\right)e^{-\frac{[k_x^2+k_y^2+(k_z-k_0)^2]}{2\Delta_k^2}}\times\\&e^{i\bold k\cdot(r\hat k_0- \bold r_0-\bold v_0 t)};
\end{split}
\end{eqnarray}
\begin{eqnarray}
\begin{split}
\gamma_2=&\int {d^3k}\left(-\frac{1}{8}\frac{k_x^2+k_y^2}{k_0^2}\right)e^{-\frac{[k_x^2+k_y^2+(k_z-k_0)^2]}{2\Delta_k^2}}\times\\
&e^{i\bold k\cdot(r\hat k_0- \bold r_0-\bold v_0 t)}
\end{split}
\end{eqnarray}
In order to calculate the above integrals, we use the trick
\begin{eqnarray}
\int dk~k~e^{-\frac{k^2}{2\Delta_k^2}}e^{-ikx}&=&i\frac{\partial}{\partial x}\int dk~e^{-\frac{k^2}{2\Delta_k^2}e^{-ikx}}\\
\int dk~k^2~e^{-\frac{k^2}{2\Delta_k^2}}e^{-ikx}&=&\frac{\partial}{\partial \left(\frac{-1}{2\Delta_k^2}\right)}\int dk~e^{-\frac{k^2}{2\Delta_k^2}}e^{-ikx}.
\end{eqnarray}
Then, we can obtain
\begin{eqnarray}
\begin{split}
\gamma_0=&(2\pi)^{3/2}\Delta_k^3e^{-\frac{(r-r_{0z}-v_0 t)^2\Delta_k^2}{2}}~e^{-\frac{b^2\Delta_k^2}{2}}\times\\&e^{ik_0(r-r_{0z}-v_0 t)};
\end{split}
\end{eqnarray}
\begin{eqnarray}
\begin{split}
\gamma_1=&(2\pi)^{3/2}\Delta_k^3\left[-\frac{1}{2k}(-ir_{0x}\Delta_k^2+r_{0y}\Delta_k^2)\right]\times\\
&e^{-\frac{(r-r_{0z}-v_0 t)^2\Delta_k^2}{2}}~e^{-\frac{b^2\Delta_k^2}{2}}~e^{ik_0(r-r_{0z}-v_0 t)};
\end{split}
\end{eqnarray}
\begin{eqnarray}
\begin{split}
\gamma_2=&(2\pi)^{3/2}\Delta_k^3\left[-\frac{1}{8k_0^2}(2\Delta_k^2-\Delta_k^4b^2)\right]\times\\
&e^{-\frac{(r-r_{0z}-v_0 t)^2\Delta_k^2}{2}}~e^{-\frac{b^2\Delta_k^2}{2}}~e^{ik_0(r-r_{0z}-v_0 t)}.
\end{split}
\end{eqnarray}
Therefore, the scattered wave can be written as
\begin{eqnarray}
\begin{split}
\psi_g^{s}(\bold r)=&-\frac{\hbar vk}{4\pi\hbar^2v^2}\frac{1}{r}\frac{1}{(2\pi)^{3/2}}\left(\frac{1}{\pi\Delta_k^2}\right)^{\frac{3}{4}}\left(\int d\bold rV(\bold r)e^{- i\bold q\cdot r}\right)\\\times
& \left(\begin{array}{cc}\rm 1+cos\theta & \rm sin\theta e^{-i\phi}\\
\rm sin\theta e^{i\phi} &\rm  1-cos\theta
\end{array}\right)~\left(\begin{array}{cc}\gamma_0+\gamma_2\\\gamma_1\end{array}\right)\\
=&\frac{1}{r}\left(\begin{array}{cc}\rm 1+cos\theta & \rm sin\theta e^{-i\phi}\\
\rm sin\theta e^{i\phi} &\rm  1-cos\theta
\end{array}\right)~\left(\begin{array}{cc}M\\N\end{array}\right),
\end{split}
\end{eqnarray}
where $M=-\frac{\hbar v k}{4 \pi\hbar^2 v^2}M_0~\Delta_k^{3/2}\pi^{-3/4}\left[1-\frac{\Delta_k^2}{4k_0^2}+\frac{\Delta_k^4}{8k_0^2}b^2\right]\times\\e^{-\frac{(r-r_{0z}-v_0 t)^2\Delta_k^2}{2}}e^{-\frac{b^2\Delta_k^2}{2}}~e^{ik_0(r-r_{0z}-v_0 t)}$ and $N=-\frac{\hbar v k}{4 \pi\hbar^2 v^2}M_0~\Delta_k^{3/2}\pi^{-3/4}\left[(-ir_{0x}\Delta_k^2+r_{0y}\Delta_k^2)\cdot \frac{1}{2k}\right]\times\\e^{-\frac{(r-r_{0z}-v_0 t)^2\Delta_k^2}{2}}e^{-\frac{b^2\Delta_k^2}{2}}~e^{ik_0(r-r_{0z}-v_0 t)}$.
\\

The probability for a detector at r in the time interval $t$---$t+dt$ to detect a particle is
\begin{eqnarray}
dP=(r^2d\Omega\times v_0 dt) \left|\psi_g^s(\bold r)\right|^2.
\end{eqnarray}
Therefore, the total probability to detect a particle in the infinite time interval can be obtained by integrating on the time $t$, i.e.,
\begin{eqnarray}
\begin{split}
P=&\int_{-\infty}^{\infty}(r^2d\Omega\times v_0 dt) |\psi_g^s|^2\\\\
=&\int_{-\infty}^{\infty}v_0dtd\Omega\left[2M^2(1+cos\theta)+\right.\\\\ &2|N|^2(1-cos\theta)+\left.Msin\theta(N^*e^{i\phi}+Ne^{-i\phi})\right]\\\\
=&\frac{2\sqrt\pi}{\Delta_k}M^2(1+cos\theta)\times\\\\&\left[1-tan{\frac{\theta}{2}}\frac{\Delta_k^2}{k_0}b~sin(\phi-\alpha)\right]~d\Omega\\\\
\approx&2(\frac{\hbar v k}{4\pi \hbar^2 v^2})^2\Delta_k^2\pi^{-1}|M_0|^2(1+cos\theta)e^{-b^2\Delta_k^2}\times\\\\
&\left[1-tan{\frac{\theta}{2}}\frac{\Delta_k^2}{k_0}b~sin(\phi-\alpha)-\frac{\Delta_k^2}{2k_0^2}\right]~d\Omega,
\end{split}
\end{eqnarray}
where $sin\alpha=b_y/b$. In the derivation of the above formula, we derive up to the order $\Delta_k^2/k_0^2$ by assuming $\Delta_k/k_0<<1$. Note that this approximation maybe not reasonable when $k_0$ is very small. But, the main point is that the full quantum calculation also gives the same physical results in the limit $\Delta_k/k_0<<1$. We can further obtain a neat form of the above formula by using the approximation
$1-tan{\frac{\theta}{2}}\frac{\Delta_k^2}{k_0}b~sin(\phi-\alpha)-\frac{\Delta_k^2}{2k_0^2}\approx exp\left\{-\left(tan{\frac{\theta}{2}}\frac{\Delta_k^2}{k_0}b~sin(\phi-\alpha)+\frac{\Delta_k^2}{2k_0^2}\right)\right\}$. Therefore, the scattering probability becomes
\begin{eqnarray}
\begin{split}
P=&2(\frac{\hbar v k}{4\pi \hbar^2 v^2})^2\Delta_k^2\pi^{-1}|M_0|^2(1+cos\theta)\times\\\\
&exp\left[-b^2\Delta_k^2-\left(tan{\frac{\theta}{2}}\frac{\Delta_k^2}{k_0}b~sin(\phi-\alpha)+\frac{\Delta_k^2}{2k_0^2}\right)\right]~d\Omega\\\\
=&2(\frac{\hbar v k}{4\pi \hbar^2 v^2})^2\Delta_k^2\pi^{-1}|M_0|^2(1+cos\theta)\times\\\\
&exp\left\{-\Delta_k^2\left[b_x^2+b_y^2+b tan\frac{\theta}{2}\frac{1}{k_0}sin(\phi-\alpha)+\frac{1}{2k_0^2}\right]\right\}~d\Omega \\\\
\approx &2(\frac{\hbar v k}{4\pi \hbar^2 v^2})^2\Delta_k^2\pi^{-1}|M_0|^2(1+cos\theta)\times\\\\
&exp\left\{-\Delta_k^2\left[\left(b_x+\frac{1}{2k_0}tan\frac{\theta}{2} sin\phi\right)^2+\right.\right.\\\\
&\left.\left.\left(b_y-\frac{1}{2k_0}tan\frac{\theta}{2} cos\phi\right)^2\right]\right\}~d\Omega \\\\
= &2(\frac{\hbar v k}{4\pi \hbar^2 v^2})^2\Delta_k^2\pi^{-1}|M_0|^2(1+cos\theta)\times\\\\
&exp\left\{-\Delta_k^2\left[\left(b_x+{\Delta_{ch}\cdot sin\phi}\right)^2+\left(b_y-{\Delta_{ch}\cdot cos\phi}\right)^2\right]\right\}~d\Omega,\\\\
\end{split}
\end{eqnarray}
where $\Delta_{ch}=\frac{1}{2k_0}tan\rm\frac{\theta}{2}$ is exact the chirality protected transverse shift as we have shown in the main text. The full quantum calculation strongly validate the semiclassical argument on how the chirality protected transverse shift leads to the anomalous scattering probability.
By definition, the plane-wave scattering cross section can be obtained by integrating the impact parameter $\bold b$, i.e.,
\begin{eqnarray}
\begin{split}
\sigma(\Omega)=&\frac{\Sigma P_i}{n}\\=&\frac{1}{n}\int n db_xdb_y\left\{2(\frac{\hbar v k}{4\pi \hbar^2 v^2})^2\Delta_k^2\pi^{-1}|M_0|^2(1+cos\theta)\right.\\\times
&\left.e^{-\Delta_k^2\left[\left(b_x+{\Delta_{ch}\cdot sin\phi}\right)^2+\left(b_y-{\Delta_{ch}\cdot cos\phi}\right)^2\right]}\right\}~d\Omega\\
=&2(\frac{\hbar v k}{4\pi \hbar^2 v^2})^2~|M_0|^2(1+cos\theta)~d\Omega,
\end{split}
\end{eqnarray}
which is fully consistent with the result obtained by plane-wave scattering.

\begin{center}
\textbf{Appendix B: The quantum lifetime and transport lifetime in Weyl Semimetals}
\end{center}
In this section, we firstly calculate the lifetime ratio $R_{\tau0}=\tau_{t0}/\tau_{q0}$ in the plane-wave scattering process.  The scattering cross section in WSMs reads as
\begin{eqnarray}
\sigma(\theta,\phi)=|f|^2=2\left(\frac{\hbar v k}{4\pi \hbar^2 v^2}\right)^2~|M_0|^2\rm(1+cos\theta).
\end{eqnarray}
In dilute impurities limit, the quantum lifetime can be obtained from the scattering cross section, i.e. $1/\tau_{q0}=v_f n_i\int \sigma(\theta)d\theta$, where $v_f$ is the Fermi velocity, $n_i$ is the impurity concentration. By contrast, the transport lifetime is $1/\tau_{t0}=v_f n_i\int\sigma(\theta){\rm (1-cos\theta)}d\theta$, where the factor $\rm (1-cos\theta)$ counts for the scattering in one direction. Thus, the lifetime ratio for plane-wave scattering turns out to be $R_{\tau 0}=2$.  As we have obtained from the main text, by considering wave-packet scattering, the quantum lifetime $\tau_q$ and transport lifetime $\tau_t$ can be defined as:
\begin{eqnarray}
\frac{1}{\tau_q}&=&v_f n_i\int_0^{b_c} d\bold b\int d\theta d\phi P(\theta,\phi,\bold b)\\
\frac{1}{\tau_{t}}&=&v_f n_i\int_0^{b_c}d\bold b \int d\theta d\phi {\rm \left(1-cos\theta\right)}P(\theta,\phi,\bold b)
\end{eqnarray}
It should be noted that the wave-packet scattering probability $P(\theta,\phi,\bold b)$ is the function of the impact parameter $\bold b$. $b_c$ is the cutoff of the impact parameter, which depends on the density of the impurities. Based on the analytic expression in Eq.(38) and Eq.(39), we calculate how the Fermi energy and the cutoff of the impact parameter affect the scattering probability. We set the width of the wave packet (k-space) as $\Delta_k=2\times10^{8}\rm m^{-1}$. Fig. 4 (a) shows the scattering probability for zero impact parameter $\bold b=0$ for different Fermi energy. Fig. 4 (b) shows how the scattering probability changes for different impact parameters. Fig. 4 (c) shows that the lifetime ratio versus the Fermi energy for different cutoff $b_c$ of the impact parameters. Notably, we renormalized the scattering probability at the scattering angle $\theta=0$, which is the demand of the Klein tunneling. We can see that the lifetime ratio decreases with the increase of the impact parameter.
\begin{figure}[!htb]
\centering
\includegraphics[height=2.8cm, width=8.8cm, angle=0]{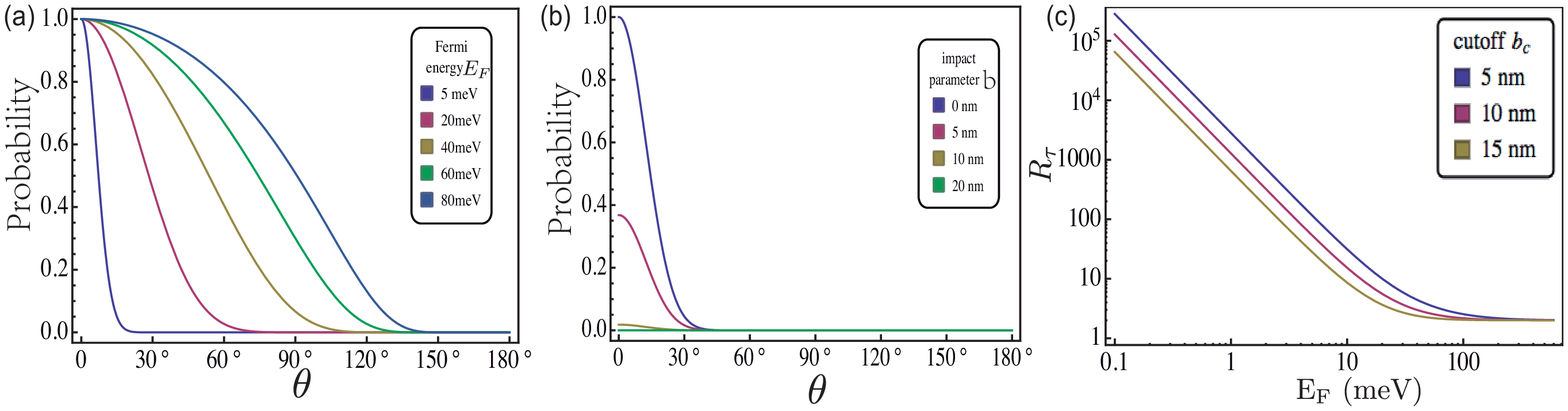}
\caption{Quantum lifetime and transport lifetime in WSMs.
(a) The scattering probability  versus the scattering angle for various Fermi energy $E_F$ in WSMs. (b) The Scattering probability versus the scattering angle for various impact parameters $\bold b$. The Fermi energy is fixed at $E_F=\rm 10~meV$, and the angle $\phi$ is fixed at $\phi=\pi/2$. (c) The lifetime ratio $R_\tau$ versus the Fermi energy for various cutoff of the impact parameters.\label{fig:S1}}
\end{figure}

\begin{center}
\textbf{Appendix C: Wave-packet Scattering for the Dirac Equation}
\end{center}

In this section, we consider both of the plane-wave scattering and wave-packet scattering for Dirac fermions in graphene.

\textit{Plane-wave scattering---}
The Hamiltonian of Dirac fermions is $\mathcal H=k_x\sigma_x+k_y\sigma_y$. The Green's function can be obtained by the Fourier transformation:\cite{dsnovikov,fguinea}
\begin{eqnarray}
G(\bold r)=-\frac{2k}{v\sqrt{2\pi i k}}\frac{e^{ikr}}{\sqrt{r}}\left(\begin{array}{cc}1 & e^{-i\phi}\\e^{i\phi} & 1\end{array}\right),
\end{eqnarray}
where $\phi$ characterizes the outgoing angle of the scattered wave.
Assume the plane wave incident in the direction $\phi=\phi_0$. Thus the incident plane wave can be written as  $\psi^{in}=\frac{1}{\sqrt{2}}\left(\begin{array}{cc}e^{-i\phi_0/2} \\e^{i\phi_0/2}\end{array}\right)e^{i\bold k\cdot \bold r}
$.
According to the first order Born approximation, the scattered wave function turns out to be
\begin{eqnarray}
\begin{split}
\psi^s(\bold r)=&\int G(\bold r-\bold r^\prime)V(\bold r^{\prime})\psi^{in}(\bold r^\prime)d\bold \bold r\\
=&-\frac{k}{v\sqrt{\pi i k}}\frac{e^{ikr}}{\sqrt{r}}\left(\begin{array}{cc}e^{-i\phi_0/2} + e^{-i(\phi-\phi_0/2)}\\e^{i(\phi-\phi_0/2)} + e^{i\phi_0/2}\end{array}\right)\times\\
&\int d\bold re^{-i\bold q\cdot \bold r}V(\bold r)\\
=&f(\bold k,\hat r)\frac{e^{ikr}}{\sqrt{r}},
\end{split}
\end{eqnarray}
where $f(\bold k,\hat r)=-\frac{k}{v\sqrt{\pi i k}}\left(\begin{array}{cc}e^{-i\phi_0/2} + e^{-i(\phi-\phi_0/2)}\\e^{i(\phi-\phi_0/2)} + e^{i\phi_0/2}\end{array}\right)\times\\\int d\bold re^{-i\bold q\cdot \bold r}V(\bold r)$ represents the scattering amplitude.
Therefore, the differential scattering cross section is
\begin{eqnarray}
\sigma(\Omega)=|f|^2=\frac{4k}{v^2\hbar^2\pi}|M_0|^2[1+cos(\phi-\phi_0)],
\end{eqnarray}
where $M_0(\bold q)=\int d\bold re^{-i\bold q\cdot \bold r}V(\bold r)$ is the scattering matrix.
At $\phi-\phi_0=\pi$(backscattering), the scattering cross section vanishes, which can be understood as the Klein tunneling for Dirac fermions.
\\\\

\textit{Wave-packet scattering---}
Using the same method as the wave-packet scattering for Schr\"odinger equation, one can obtain the general outgoing wavefunction for Dirac fermions, i.e.,
\begin{eqnarray}
\begin{split}
\psi_g(\bold r,t)=&\int \frac{d^2k}{(2\pi)}\varphi(\bold k)e^{-i\bold k\cdot \bold r_0-i Et}
\left[\frac{1}{\sqrt{2}}\left(\begin{array}{cc}e^{-i\phi_0/2}\\e^{i\phi_0/2}\end{array}\right)-\right.\\\\
&\left.\frac{2k}{v\sqrt{2\pi i k}}\frac{e^{ikr}}{\sqrt{r}}\left(\begin{array}{cc}1 & e^{-i\phi}\\e^{i\phi} & 1\end{array}\right)~\frac{1}{\sqrt{2}}\left(\begin{array}{cc}e^{-i\phi_0/2}\\e^{i\phi_0/2}\end{array}\right)\times\right.\\\\&\left.
\int d\bold rV(\bold r)e^{- i\bold q\cdot r}\right]\\\\
\approx &\int \frac{d^2k}{(2\pi)}\varphi(\bold k)e^{-i\bold k\cdot(\bold r_0+\bold v_0 t)}
\left[\frac{1}{\sqrt{2}}\left(\begin{array}{cc}e^{-i\phi_0/2}\\e^{i\phi_0/2}\end{array}\right)e^{i\bold k\cdot\bold r}-\right.\\\\
&\left.\frac{2k}{v\sqrt{2\pi i k}}\frac{e^{ikr}}{\sqrt{r}}\left(\begin{array}{cc}1 & e^{-i\phi}\\e^{i\phi} & 1\end{array}\right)~\frac{1}{\sqrt{2}}\left(\begin{array}{cc}e^{-i\phi_0/2}\\e^{i\phi_0/2}\end{array}\right)\right.\times\\\\
&\left.\int d\bold rV(\bold r)e^{- i\bold q\cdot r}\right]\\\\
\approx &\int \frac{d^2k}{(2\pi)}\varphi(\bold k)e^{-i\bold k\cdot(\bold r_0+\bold v_0 t-\bold r)}
\left[\frac{1}{\sqrt{2}}\left(\begin{array}{cc}e^{-i\phi_0/2}\\e^{i\phi_0/2}\end{array}\right)\right]\\\\-&
\frac{2k_0}{v\sqrt{2\pi i k_0}}\frac{M_0}{\sqrt{r}}\left(\begin{array}{cc}e^{-i\phi_0/2}\\e^{i\phi_0/2}\end{array}\right)\int \frac{d^2k}{(2\pi)}\varphi(\bold k)\times\\\\
&e^{i[kr-\bold k\cdot(\bold r_0+\bold v_0 t)]}
\left[\frac{1}{\sqrt{2}}\left(\begin{array}{cc}e^{-i\phi_0/2}\\e^{i\phi_0/2}\end{array}\right)\right],
\end{split}
\end{eqnarray}
where $\varphi(\bold k)=\left(\frac{1}{\pi\Delta_k^2}\right)^{\frac{1}{2}}e^{\frac{-(\bold k-\bold k_0)^2}{2\Delta_k^2}}$ is a 2D Gaussian distribution function, $\bold r_0$ is the initial position of the wave packet, $\bold v_0$ is the velocity of the incident wave packet,  $M_0=\int d\bold r~e^{-i\bold q\cdot \bold r}V(\bold r)$ is the scattering matrix, and $\phi_0(\bold k)=arcsin\left(\frac{k_y}{k}\right)$. Notice we have used following approximations in the derivation of the above formulas:
\\\\
(1) $E=v\sqrt{k_x^2+k_y^2+k_z^2}\approx v\sqrt{k_{x0}^2+k_{y0}^2+k_{z0}^2}+\frac{\partial E}{\partial k_x}|_{k_x=k_{x0}}(k_x-k_{x0})+\frac{\partial E}{\partial k_y}|_{k_y=k_{y0}}(k_y-k_{y0})+\frac{\partial E}{\partial k_z}|_{k_z=k_{z0}}(k_z-k_{z0})=vk_0+v\hat k_0\cdot(\bold k-\bold k_0)=v\hat k_0\cdot\bold k=\bold v_0\cdot \bold k$.
\\\\
(2) $M_0(\bold q)=M_0(k\hat r-\bold k)\approx M_0(k\hat r-\bold k_{0})$. That's why we can take $M_0$ out of the integral. Actually, if we consider a delta potential, then $M_0$ is a constant. So we can always take $M_0$ out of the integral for a delta potential.  For a general potential, taking $M_0$ out of the integral $\int d^3 k$ is actually an approximation.\\
\\
(3) Since $f(\bold k,\hat r)=-\frac{k}{v\sqrt{\pi i k}}\left(\begin{array}{cc}e^{-i\phi_0/2} + e^{-i(\phi-\phi_0/2)}\\e^{i(\phi-\phi_0/2)} + e^{i\phi_0/2}\end{array}\right)\times\\
\int d\bold re^{-i\bold q\cdot \bold r}V(\bold r)$, we expand $f(\bold k,\hat r)$ around $\bold k_0$. Since $f(\bold k)\propto \sqrt{k}$, we expand $\sqrt{k}$ around $\bold k_0$ first. By Taylor expansion, we find that $\sqrt{k}=(k_x^2+k_y^2+k_z^2)^{\frac{1}{4}}\approx \frac{k_0^2+k_0(\hat k_0\cdot \bold k)}{2 k_0^{\frac{3}{2}}}\approx \sqrt{k_0}$ (to the first order approximation).  $\phi_0$ is the incident angle of the plane wave, and we can expand $\phi_0$ near $\phi_0=0$ (x-axis). Thus, $e^{i\phi_0/2}\approx1+i\phi_0\approx 1+ i\frac{k_y}{2k_0}$ and $e^{-i\phi_0/2}\approx1-i\phi_0\approx 1- i\frac{k_y}{2k_0}$.\\
The scattered wavefunction then can be inferred from Eq.(43):
\begin{eqnarray}
\begin{split}
\psi_g^s=&-\frac{2k_0}{v\sqrt{2\pi i k_0}}\frac{M_0}{\sqrt{r}}\left(\begin{array}{cc}1 & e^{-i\phi}\\e^{i\phi} & 1\end{array}\right)\times\\
&
\int \frac{d^2k}{(2\pi)}\varphi(\bold k)e^{i[kr-\bold k\cdot(\bold r_0+\bold v_0 t)]}
\left[\frac{1}{\sqrt{2}}\left(\begin{array}{cc}e^{-i\phi_0/2}\\e^{i\phi_0/2}\end{array}\right)\right]\\
=&\left(\begin{array}{cc}1 & e^{-i\phi}\\e^{i\phi} & 1\end{array}\right)\left(\begin{array}{cc} M-N/2\\M+N/2\end{array}\right)\frac{1}{\sqrt{r}},
\end{split}
\end{eqnarray}
where
$M=-\frac{k_0}{v\sqrt{\pi i k_0}}M_0~\int \frac{d^2k}{(2\pi)}\varphi(\bold k)e^{i[kr-\bold k\cdot(\bold r_0+\bold v_0 t)]}
$, and
$N=-\frac{k_0}{v\sqrt{\pi i k_0}}M_0~\int \frac{d^2k}{(2\pi)}\varphi(\bold k)(i\frac{k_y}{k})e^{i[kr-\bold k\cdot(\bold r_0+\bold v_0 t)]}$. Indeed, we can use the relation $N=-\frac{1}{k}\frac{\partial M}{\partial r_{0y}}$ to obtain $N$.
The probability for a detector at r in the time interval $t$---$t+dt$ to detect a particle is
\begin{eqnarray}
dP=(r^2d\Omega\times v_0 dt) \left|\psi_g^s(\bold r)\right|^2.
\end{eqnarray}
Thus, the total probability to detect a particle in the infinite time interval can be obtained by integrating on the time $t$, i.e.,
\begin{eqnarray}
\begin{split}
P=&\int_{-\infty}^{\infty}(r~d\Omega\times v_0 dt) |\psi_g^s|^2\\
=&\int_{-\infty}^{\infty}( d\Omega\times v_0 dt) \left[4 M^2\left(1+cos\phi\right)+\right.\\
&\left.N^2\left(1-cos\phi\right)\right]\\
\approx&\frac{4k_0}{\pi v^2\hbar^2}|M_0|^2~(1+cos\phi) {{\frac{\Delta_k}{\sqrt{\pi}}e^{-b^2\Delta_k^2}}}~d\Omega,
\end{split}
\end{eqnarray}
where $b$ is the impact parameter of the wave packet. In the derivation of the above formula, we derive up to the first order of $N$ by ignoring the term $N^2$, because $N^2\approx(\Delta_k^2/k_0)^2<<1$. This result also shows that the scattering probability is exponentially decay with the increasing of the impact parameter.
The plane-wave scattering cross section then can be obtained from the wave-packet scattering probability:
\begin{eqnarray}
\begin{split}
\sigma(\Omega)=&\frac{\Sigma P_i}{n}\\=&\frac{\int_{-\infty}^{\infty}db\left[n\cdot\frac{4k_0}{\pi v^2\hbar^2}|M_0|^2~{e^{-\Delta_k^2 b^2}}~(1+cos\phi)~\right]}{n}\\
=&\frac{4k_0}{v^2\hbar^2\pi}|M_0|^2 (1+cos\phi)~d\Omega.
\end{split}
\end{eqnarray}

\begin{center}
\textbf{Appendix D: Calculation on the center of the incident wave packet}
\end{center}

Since the eigenfunction for Dirac equation and Weyl equation is in the spinor form, we need to identify the center of the incident wave-packet in graphene and Weyl semimetals.

\textit{Center of the incident wave packet in graphene.---}
The incident wave packet in graphene is
\begin{eqnarray}
\begin{split}
\psi^{in}_{g}=&\int \frac{d^2k}{(2\pi)}\varphi(\bold k)e^{-i\bold k\cdot(\bold r_0+\bold v_0 t-\bold r)}
\left[\frac{1}{\sqrt{2}}\left(\begin{array}{cc}e^{-i\phi_0/2}\\e^{i\phi_0/2}\end{array}\right)\right]\\&=\left(\begin{array}{cc}C_1\\C_2\end{array}\right)
\end{split}
\end{eqnarray}
Now, we obtain the first spinor and the second spinor, respectively.

The first spinor is:
\begin{eqnarray}
\begin{split}
C_1=&\frac{1}{\sqrt{2}}\int \frac{d^2k}{(2\pi)}\varphi(\bold k)e^{-i\bold k\cdot(\bold r_0+\bold v_0 t-\bold r)}e^{-i\phi_0/2}\\
=& \frac{1}{\sqrt{2}}\int \frac{d^2k}{(2\pi)}\left(\frac{1}{\pi\Delta_k^2}\right)^{\frac{1}{2}}e^{-\frac{(\bold k-\bold k_0)^2}{2\Delta_k^2}}e^{-i\bold k\cdot(\bold r_0+\bold v_0 t-\bold r)}e^{-i\phi_0/2}\\
=&  \frac{1}{\sqrt{2}}\frac{d^2k}{(2\pi)}\left(\frac{1}{\pi\Delta_k^2}\right)^{\frac{1}{2}}\int dk_xdk_y~e^{-\frac{(\bold k-\bold k_0)^2}{2\Delta_k^2}}\times\\
&e^{-i\bold k\cdot(\bold r_0+\bold v_0 t-\bold r)}(1-i\frac{k_y}{2k_0})\\
=&C-C_\Delta,
\end{split}
\end{eqnarray}
where
\begin{eqnarray}
\begin{split}
C=&\frac{1}{\sqrt{2}}\int \frac{d^2k}{(2\pi)}\varphi(\bold k)e^{-i\bold k\cdot(\bold r_0+\bold v_0 t-\bold r)}\\
=& \frac{1}{\sqrt{2}}\int \frac{d^2k}{(2\pi)}\left(\frac{1}{\pi\Delta_k^2}\right)^{\frac{1}{2}}e^{-\frac{(\bold k-\bold k_0)^2}{2\Delta_k^2}}e^{-i\bold k\cdot(\bold r_0+\bold v_0 t-\bold r)}\\
=&  \frac{1}{\sqrt{2}}\frac{1}{(2\pi)}\left(\frac{1}{\pi\Delta_k^2}\right)^{\frac{1}{2}}\int dk_xdk_y~e^{-\frac{(\bold k-\bold k_0)^2}{2\Delta_k^2}}e^{-i\bold k\cdot(\bold r_0+\bold v_0 t-\bold r)}\\
=& \frac{1}{\sqrt{2}}\frac{1}{(2\pi)}\left(\frac{1}{\pi\Delta_k^2}\right)^{\frac{1}{2}} \left(\int dk_xe^{-\frac{(k_x-k_{x0})^2}{2\Delta_k^2}e^{-ik_x(x_0+v_0t-x)}}\right)\\
&\times\left(\int dk_y~e^{-\frac{(k_y-k_{y0})^2}{2\Delta_k^2}e^{-ik_y(y_0-y)}}\right)\\
=& \frac{1}{\sqrt{2}}\frac{1}{(2\pi)}\left(\frac{1}{\pi\Delta_k^2}\right)^{\frac{1}{2}}
\left(\sqrt{2\pi\Delta_k^2}e^{-\frac{(x-x_0-v_0t)^2\Delta_k^2}{2}}\right) \times\\
&\left(\sqrt{2\pi\Delta_k^2}e^{-\frac{(y-y_0)^2\Delta_k^2}{2}}\right)~e^{-ik_{x0}(x-x_0-v_0t)}\\
=&\frac{\Delta_k}{\sqrt{2\pi}}e^{-\frac{(x-x_0-v_0t)^2\Delta_k^2}{2}}e^{-\frac{(y-y_0)^2\Delta_k^2}{2}}~e^{-ik_{x0}(x-x_0-v_0t)}
\end{split}
\end{eqnarray}
and
\begin{eqnarray}
\begin{split}
C_{\Delta}=&\frac{1}{\sqrt{2}}\frac{d^2k}{(2\pi)}\left(\frac{1}{\pi\Delta_k^2}\right)^{\frac{1}{2}}\int dk_xdk_y~e^{-\frac{(\bold k-\bold k_0)^2}{2\Delta_k^2}}\times\\
&e^{-i\bold k\cdot(\bold r_0+\bold v_0 t-\bold r)}(i\frac{k_y}{2k_0})\\
=& \frac{1}{\sqrt{2}}\frac{d^2k}{(2\pi)}\left(\frac{1}{\pi\Delta_k^2}\right)^{\frac{1}{2}} \frac{i}{2k_0}\times\\
&\left(\int dk_xe^{-\frac{(k_x-k_{x0})^2}{2\Delta_k^2}e^{-ik_x(x_0+v_0t-x)}}\right)\times\\
&\left(\int dk_y~e^{-\frac{(k_y-k_{y0})^2}{2\Delta_k^2}e^{-ik_y(y_0-y)}}k_y\right)\\
=& \frac{1}{\sqrt{2}}\frac{d^2k}{(2\pi)}\left(\frac{1}{\pi\Delta_k^2}\right)^{\frac{1}{2}}\frac{i}{2k_0} \left(\sqrt{2\pi\Delta_k^2}e^{-\frac{(x-x_0-v_0t)^2\Delta_k^2}{2}}\right.\times\\
&\left.e^{-ik_{x0}(x-x_0-v_0t)}\right) \left(i\frac{\partial}{\partial y_0}\right)\left(\sqrt{2\pi\Delta_k^2}e^{-\frac{(y-y_0)^2\Delta_k^2}{2}}\right)\\
=&\frac{\Delta_k}{\sqrt{2\pi}}\left(-\frac{(y-y_0)\Delta_k^2}{2k_0}\right)e^{-\frac{(x-x_0-v_0t)^2\Delta_k^2}{2}}e^{-\frac{(y-y_0)^2\Delta_k^2}{2}}\times\\
&e^{-ik_{x0}(x-x_0-v_0t)}
\end{split}
\end{eqnarray}
Therefore, $C_1=\left(1+\frac{(y-y_0)\Delta_k^2}{2k_0}\right)C$.
In the same way, we can obtain the second spinor, which is
\begin{eqnarray}
\begin{split}
C_2=&\frac{1}{\sqrt{2}}\int \frac{d^2k}{(2\pi)}\varphi(\bold k)e^{-i\bold k\cdot(\bold r_0+\bold v_0 t-\bold r)}e^{i\phi_0/2}\\
=& \frac{1}{\sqrt{2}}\int \frac{d^2k}{(2\pi)}\left(\frac{1}{\pi\Delta_k^2}\right)^{\frac{1}{2}}e^{-\frac{(\bold k-\bold k_0)^2}{2\Delta_k^2}}e^{-i\bold k\cdot(\bold r_0+\bold v_0 t-\bold r)}e^{i\phi_0/2}\\
=&  \frac{1}{\sqrt{2}}\frac{d^2k}{(2\pi)}\left(\frac{1}{\pi\Delta_k^2}\right)^{\frac{1}{2}}\int dk_xdk_y~e^{-\frac{(\bold k-\bold k_0)^2}{2\Delta_k^2}}\times\\
&e^{-i\bold k\cdot(\bold r_0+\bold v_0 t-\bold r)}(1+i\frac{k_y}{2k_0})\\
=&C+ C_\Delta.
\end{split}
\end{eqnarray}
Therefore, the center of the wave packet is
\begin{eqnarray}
\begin{split}
\bar y=&\langle\psi_g^{in}|y|\psi_g^{in}\rangle\\
=&\int dyd x \left(\begin{array}{cc}C_1^*& C_2^*\end{array}\right) y \left(\begin{array}{cc}C_1\\ C_2\end{array}\right)\\
=& \int dyd x \left(C_1^2 y+C_2^2 y\right)\\
=& \frac{1}{2}\left(y_0+\frac{1}{2k_0}\right)+\frac{1}{2}\left(y_0-\frac{1}{2k_0}\right)\\
=& y_0
\end{split}
\end{eqnarray}
Therefore, the impact parameter in graphene is $b= y_0$. Hence, the scattering probability
$P\propto e^{-b^2}$, which indicates that the scattering probability is exponentially decay with the increasing of the impact parameter.

\textit{Center of the incident wave packet in WSMs.---}
The incident wave packet is
\begin{eqnarray}
\begin{split}
\psi^{in}_{g}=&\int \frac{d^3k}{(2\pi)^{3/2}}\varphi(\bold k)e^{-i\bold k\cdot(\bold r_0+\bold v_0 t-\bold r)}
\left(\begin{array}{cc}cos\frac{\theta_0}{2}\\sin\frac{\theta_0}{2}e^{i\phi_0}\end{array}\right)\\
=&\left(\begin{array}{cc}S_1\\S_2\end{array}\right)
\end{split}
\end{eqnarray}
We calculate the first spinor $S_1$ and the second spinor $S_2$ in the following, respectively.

The first spinor is:
\begin{eqnarray}
\begin{split}
S_1=&\int \frac{d^3k}{(2\pi)^{3/2}}\varphi(\bold k)e^{-i\bold k\cdot(\bold r_0+\bold v_0 t-\bold r)}\left(1-\frac{k_x^2+k_y^2}{8 k_0^2}\right)\\
=&\Delta_k^{\frac{3}{2}}\pi^{-\frac{3}{4}}\left\{ 1+\frac{\Delta_k^4}{8 k_0^2}\left[(x-x_0)^2+(y-y_0)^2\right]\right\}\times\\
&e^{-\frac{(z-z_0-v_0t)^2\Delta_k^2}{2}}e^{-\frac{(y-y_0)^2\Delta_k^2}{2}}e^{-\frac{(x-x_0)^2\Delta_k^2}{2}}\times\\
&e^{-ik_{z0}(z-z_0-v_0t)}\\
\approx &\Delta_k^{\frac{3}{2}}\pi^{-\frac{3}{4}}e^{-\frac{(z-z_0-v_0t)^2\Delta_k^2}{2}}e^{-\frac{(y-y_0)^2\Delta_k^2}{2}}e^{-\frac{(x-x_0)^2\Delta_k^2}{2}}\\
\times&e^{-ik_{z0}(z-z_0-v_0t)}
\end{split}
\end{eqnarray}

The second spinor is:
\begin{eqnarray}
\begin{split}
S_2=&\int \frac{d^3k}{(2\pi)^{3/2}}\varphi(\bold k)e^{-i\bold k\cdot(\bold r_0+\bold v_0 t-\bold r)}\left(\frac{k_r}{2 k_z}\right)e^{i\phi_0}\\
\approx &\int \frac{d^3k}{(2\pi)^{3/2}}\varphi(\bold k)e^{-i\bold k\cdot(\bold r_0+\bold v_0 t-\bold r)}\left(\frac{k_x+ik_y}{2k_0}\right)\\
=&\Delta_k^{\frac{3}{2}}\pi^{-\frac{3}{4}}e^{-\frac{(z-z_0-v_0t)^2\Delta_k^2}{2}}~e^{-ik_{z0}(z-z_0-v_0t)}\times\\
&\left(i\frac{\partial }{\partial x_0}-\frac{\partial }{\partial y_0}\right)\frac{1}{2k}\left[e^{-\frac{(y-y_0)^2\Delta_k^2}{2}}e^{-\frac{(x-x_0)^2\Delta_k^2}{2}}\right]\\
=&\Delta_k^{\frac{3}{2}}\pi^{-\frac{3}{4}}e^{-\frac{(z-z_0-v_0t)^2\Delta_k^2}{2}}~e^{-ik_{z0}(z-z_0-v_0t)}\times\\
&\left[i(x-x_0)-(y-y_0)\right]\times\\
&\frac{\Delta_k^2}{2k_0}
\left[e^{-\frac{(y-y_0)^2\Delta_k^2}{2}}e^{-\frac{(x-x_0)^2\Delta_k^2}{2}}\right]
\end{split}
\end{eqnarray}
The center of the wave packet is at $(\bar x, \bar y)$, where
\begin{eqnarray}
\begin{split}
\bar y=&\langle\psi_g^{in}|y|\psi_g^{in}\rangle\\
=&\int dxdyd z \left(\begin{array}{cc}S_1^*& S_2^*\end{array}\right) y \left(\begin{array}{cc}S_1\\ S_2\end{array}\right)\\
=& y_0+\frac{\Delta_k^2}{4k_0^2}y_0\\
\approx & y_0
\end{split}
\end{eqnarray}
and
\begin{eqnarray}
\begin{split}
\bar x=&\langle\psi_g^{in}|x|\psi_g^{in}\rangle\\
=&\int dxdyd z \left(\begin{array}{cc}S_1^*& S_2^*\end{array}\right) x \left(\begin{array}{cc}S_1\\ S_2\end{array}\right)\\
=& x_0+\frac{\Delta_k^2}{4k_0^2}x_0\\
\approx & x_0.
\end{split}
\end{eqnarray}
Therefore, the incident wave packet center is still at $(x_0, y_0)$, which means the impact parameter is $b=\sqrt{x_0^2+y_0^2}$. However, the chirality protected transverse shift can further modify the impact parameter, which leads to the anomalous scattering probability in WSMs.

\hspace{3mm}


\begin{thebibliography}{}
\bibitem{footnote2}
The saturation value of the lifetime ratio is $R_{\tau 0}=2$, which is the exact results calculated from plane-wave scattering process.[see the Appendix]
\end{thebibliography}
\end{document}